\documentclass{article}

\usepackage{arxiv}

\usepackage{hyperref}
\hypersetup{
    colorlinks=true,
    citecolor=black,
    linkcolor=black,
    urlcolor=blue
}
\usepackage[round]{natbib}
\usepackage[utf8]{inputenc} 
\usepackage[T1]{fontenc}    
\usepackage{url}            
\usepackage{booktabs}       
\usepackage{amsfonts}       
\usepackage{nicefrac}       
\usepackage{microtype}      
\usepackage{lipsum}		
\usepackage{graphicx}
\usepackage{doi}
\usepackage{amsmath,amssymb}
\usepackage{longtable}
\usepackage{multirow}
\usepackage{multicol}
\usepackage{subcaption}
\usepackage{array}
\usepackage{placeins}
\usepackage{tikz}
\usetikzlibrary{calc,positioning}
\usepackage{pdflscape}
\usepackage{cancel}
\usepackage{extpfeil}

\title{Towards a Theory of Quantitative Vulnerability Analysis for Engineering Systems}

\date{} 					

\author{Daniel Eisenberg \\
	Department of Operations Research\\
	Naval Postgraduate School\\
	Monterey, CA, USA \\
	\texttt{daniel.eisenberg@nps.edu} \\
    \And
	David Alderson \\
	Department of Operations Research\\
	Naval Postgraduate School\\
	Monterey, CA, USA \\
	\texttt{dlalders@nps.edu} \\
}

\renewcommand{\headeright}{Eisenberg \& Alderson 2026}
\renewcommand{\undertitle}{Last Updated: \today}
\renewcommand{\shorttitle}{Theory of Quantitative Vulnerability Analysis}

\hypersetup{
pdftitle={runtitle},
pdfsubject={q-bio.NC, q-bio.QM},
pdfauthor={Eisenberg and Alderson},
pdfkeywords={keywords},
}

\begin{document}
\maketitle

\begin{abstract}
Since the 1980s, risk analysis for engineering systems has been defined based on a tuple of scenarios, $\mathcal{S}$, likelihoods, $L$, and consequences, $C$, where $Risk = (\mathcal{S},L,C)$. We argue that equating this triplet with risk alone is insufficient and potentially misleading. Instead, we define this tuple as a vulnerability space $\mathcal{V} = (\mathcal{S},L,C)$, which becomes risk only when one makes specific interpretations of likelihood and consequence. Here, an \textit{interpretation} has mathematical meaning, and defines a map that assigns data elements about system vulnerability to a \textit{perspective}. Formally, we define a vulnerability \textit{perspective} as a specific choice of interpretations of likelihood and consequence data that allows vulnerability analysis measures to be defined and analyzed. We argue that risk is only one of four common perspectives on $\mathcal{V}$ and demonstrate mathematically why each perspective should remain a distinct concept. First, we show how interpretation of $C$ as a set-valued function produces reliability measures, not risk measures. Then, we show how interpretation of $L$ via possibility theory will produce adversarial and safety measures distinct from risk. This new theoretical framework clarifies the conceptual, mathematical, and practical differences between risk, reliability, adversarial, and safety perspectives---four common engineering system analyses often conflated in the literature. A consequence of our theory is that no single perspective will reveal all vulnerabilities of an engineering system, suggesting decision-making via risk analysis may incur unavoidable and underappreciated tradeoffs from other vulnerability perspectives. 
\end{abstract}

%
%








\section{Introduction}\label{sec:intro}

This work is concerned with how to define and apply quantitative vulnerability analysis in an engineering context. We define the vulnerability of an engineering system in terms of its susceptibility to undesirable change. This definition aligns linguistically with well-established definitions for social, human, and environmental vulnerability \citep{TurnerEtAl2003} and canonical work on engineering system resilience \citep{bruneau2003framework,bruneau2006overview,cutter2008place,mentges2023resilience}. However, such definitions provide little guidance on how to quantitatively measure vulnerability of a given engineering system or to improve its resilience with new design, leading to numerous studies applying the generic concept with disjoint and divergent mathematical treatment \citep{mentges2023resilience}. The purpose of this work is to link mathematical methods and system concepts for more practical use. 

The most prevalent quantitative approach to describe the vulnerability of a system is risk analysis. The original quantitative definition of risk is often attributed to \citet{kaplan1981quantitative} and involves analysis of three data elements:

\begin{itemize}
    \item Initiating events or scenarios, $s \in \mathcal{S}$, that define ``what can go wrong''; 
    \item Likelihoods, $L:\mathcal{S} \rightarrow [0,1]$, that measure how likely it is for a given scenario to happen; and,
    \item Consequences, $C:\mathcal{S} \rightarrow \mathbb{R}_{\geq 0}$, that measure how much the system changes for the worse as a result of the scenario.\footnote{We restrict our consequence measures to be positive, but this is simply for convenience. Our arguments hold for any consequence measure in $\mathbb{R}$.}
\end{itemize}

This definition has led many authors to define risk as a composite of these three elements, $Risk = (\mathcal{S},L,C)$, and use this definition as a starting point for analysis \citep{kaplan1981quantitative}. This definition supported decades of research that has improved our understanding of how to apply risk analysis and risk management to diverse engineering contexts \citep{aven2015risk,aven2019science,aven2024risk}. 

We argue that equating this triplet, without qualification, to the definition of risk is incorrect and potentially misleading. Instead, an unambiguous notion of risk follows only when one takes specific \textit{interpretations} of likelihood and consequence. We argue that this tuple more appropriately defines a \textit{vulnerability space}, $\mathcal{V} = (\mathcal{S},L,C)$, over which interpretations of $L$ and $C$ produce at least four distinct \textit{perspectives} on engineering system vulnerability, only one of which is risk. Here, a \textit{perspective} is a specific combination of interpretations for $L$ and $C$ that, in turn, form a metric-measure space where vulnerability can be assessed and potential system improvements prioritized.

Our new approach to define vulnerability, risk, and related concepts has significant conceptual and practical implications for engineering systems. One consequence of our work is the necessary reframing of the conceptual vocabulary for quantitative analysis of engineering systems. For example, our work argues that risk analysis is a sub-concept of system vulnerability, and is more comparable to other forms of analysis like reliability, rather than broader systems concepts like resilience \citep{cutter2008place}. This would suggest that significant research attempting to treat risk, reliability, and resilience as interchangeable and directly comparable concepts made a category mistake --- risk and reliability are specific perspectives used to transform data on state likelihoods and consequences into a system evaluation criteria, whereas resilience is a family of decision-evaluation perspectives for assessing how alternative engineering designs, policies, or configurations change system performance under disturbance (c.f., \citealt{park2013integrating,linkov2018tiered,aven2022some,logan2022risk,Aven2026}). This reframing does not necessarily obviate past comparative work but does suggest that future research should focus more on understanding the differences among vulnerability perspectives and their impact on decisions, rather than between a given vulnerability perspective and resilience.

A key insight is that the act of interpretation and forming a perspective simultaneously reveals and obscures information about engineering system vulnerability. The nature of analysis as both enabling and narrowing one's perspective on a system has been broadly established in resilience engineering theory, but without a fundamental mathematical basis \citep{woods2010capturing,woods2018theory}. We show this quantitatively: without interpretation of $L$ and $C$, there simply is not sufficient mathematical detail to create a quantitative vulnerability measure useful for decision-making. Likewise, there is no single perspective on $L$ and $C$ that is sufficient to reveal all vulnerabilities of a system for decision-making. Moreover, different perspectives often point to different vulnerabilities, even in the same system, resulting in contradictory recommendations and unavoidable tradeoffs in decision-making.

The contributions of this work as summarized as follows:
\begin{enumerate}
    \item We redefine well-established engineering system analysis methods of risk, reliability, adversarial vulnerability, and safety as perspectives on a vulnerability space.
    \item We explain the differences in interpretation of $\mathcal{V}$ that produce an formal mathematical perspective for each.
    \item We provide examples that demonstrate how different perspectives lead to distinct and often conflicting vulnerability analysis and decision-making for the same system.
    \item We further explain how choice of perspective influences analytical effort to make a decision, which may produce ``blind spots'' when choosing one perspective over another.
    \item We conclude with a discussion of multiobjective and hybrid measures that seemingly integrate perspectives but do not address fundamental distinctions and issues when trying to make a comprehensive vulnerability analysis decision.
\end{enumerate}

\section{Risk as a Perspective on Vulnerability}

If we assume risk encompasses all  $(\mathcal{S},L,C)$, then its basic definition is too vague to perform risk analysis. The tuple requires additional information to relate $L$ and $C$ meaningfully and to evaluate if one undesirable system state is worse or better than other. The majority of risk analysis research implicitly interprets $L$ and $C$ as a natural step in analysis without explanation as to why \citep{kaplan1981quantitative,aven2011risk,zio2025advances}. The reason is because $\mathcal{V}$ is currently missing necessary information about how to combine, normalize, or compare two likelihoods. Consider two scenarios $s_1, s_2 \in \mathcal{S}$ with associated likelihoods $L(s_{1})$ and $L(s_{2})$. In general, it is unknown if $L(s_{1}) \oplus L(s_{2})$ is non-idempotent to produce $L(s_{1}) + L(s_{2})$ (normal addition) or $L(s_{1}) \cdot L(s_{2})$ (normal multiplication), or if it represents an idempotent operation like $\min(L(s_{1}), L(s_{2}))$ or $\max(L(s_{1}), L(s_{2}))$. For the same reason, $\mathcal{V}$ also does not encode enough information on how to combine, normalize, or compare two consequences $C(s_1)$ and $C(s_2)$, as their relative importance depends on $L$.

Instead, risk analysis requires taking a perspective on $L$ and $C$. The most common perspective assigned by the risk analysis community is that $L$ is a probability of a given scenario and $C$ is a scalar numeric value for each scenario. While the \citet{kaplan1981quantitative} definitions of $L$ and $C$ above may appear to already achieve these goals, they do not. Risk analysis requires interpretations, denoted here as $P(\mathcal{S})$ and $G(\mathcal{S})$ respectfully,  defined as follows.

\begin{itemize}
    \item \textit{Likelihood Interpretation as Probability:} To interpret $L$ as a probability, the map $P$ is an operator on the entire function space $L$ to ensure it meets the basic definitions of a probability space, i.e.,  $P:$ $[0,1]^{|\mathcal{S}|} \rightarrow P(\mathcal{S})$ where $[0,1]^{|\mathcal{S}|}$ is the set of state likelihoods $\{L(s) \mid s \in \mathcal{S}\}$ and $P(\mathcal{S})$ ensures proper normalization of likelihoods to probabilities $ = \{p:\mathcal{S} \rightarrow [0,1] \mid \sum_{s \in \mathcal{S}}{p(s) = 1} \}$. Assuming a standard discrete normalization, then $p(s) = \frac{L(s)}{\sum_{s'\in \mathcal{S}}{L(s')}}$.
    
    \item \textit{Consequence Interpretation as a Numeric Value:} To interpret $C$ as a number, we define $G$ as a point-wise function composition that converts consequence data $C(s)$ into a real-valued function over the state space. We define  $G:C(s)\rightarrow \mathbb{R}$, so if $G(s_{1}) > G(s_{2})$, then state $s_{1}$ is more consequential than $s_{2}$. $G$ only induces a partial ordering, as multiple states may have the same consequence: $s_{1} \neq s_{2}$, and $G(s_{1}) = G(s_{2})$. This means $G$ is not a metric by mathematical definition as it assigns a scalar value to each state, rather than a distance, but it can be defined by a psuedometric distance $d_{G}(s,t) = |G(s) - G(t)|$ that satisfies non-negativity, symmetry, and the triangle inequality. For convenience, we assume interpretation involves scaling, such that $G: C(s) \rightarrow [0,1]$ where $0$ means no consequence and $1$ is maximum consequence over $\mathcal{S}$.
\end{itemize} 

We define risk as a perspective on $\mathcal{V}$, specifically $Risk(\mathcal{V}) = (\mathcal{S},P(\mathcal{S}),G(\mathcal{S}))$, with $P$ for probability and $G$ for a scalar function. In what follows, we often abbreviate $Risk = (P,G)$.

With this definition, we now have a powerful basis to quantify the vulnerability of a system via any risk measure that combines $P$ with $G$. This includes simply estimating the expected consequences of a given scenario, $\mathbb{E}[G(s)] = P(s) \cdot G(s)$, total expect loss over all scenarios, $\mathbb{E}[G(\mathcal{S})] = \frac{1}{|\mathcal{S}|}\sum_{s \in \mathcal{S}}{P(s) \cdot G(s)}$, or any relevant measure that is weighted via the distribution of scenario probabilities, e.g., cumulative value at risk ($CVaR$) \citep{tyrrell2015engineering,norton2021calculating,chaudhuri2022certifiable}. Any and all of these techniques fall within this risk perspective on $\mathcal{V}$. Risk analysis involves choosing an evaluative function after taking the $(P,G)$ perspective and using it to compare and prioritize vulnerabilities. 

Defining risk as a vulnerability perspective ensures that all associated risk measures remain coherent and convex \citep{artzner1999coherent,rockafellar2007coherent, rockafellar2000optimization}, which are qualities required to ensure effective decision-making. Importantly, these features are independent of the nature or derivation of the probabilities in a risk measure itself. If $\mathbb{E}[G(s_{1})] > \mathbb{E}[G(s_{2})]$, then scenario $s_{1}$ will have a greater effect on risk-based decisions irrespective if $P(s)$ is derived from frequentist, Bayesian, or otherwise statistical techniques. Likewise, the interpretation of the consequences in a quantitative risk measure does not affect decision-making use as long as worse scenarios receive worse values. If $G$ is interpreted in dollars, lives, or any other countable and valued measure, it will still form a risk measure if combined with a $P$.


\section{Reliability as a Perspective on Vulnerability}

To strengthen our argument that risk is a perspective on vulnerability, we show that  reliability analysis is also a perspective on vulnerability. Assuming risk is equivalent to $\mathcal{V}$ would conflate reliability measures as risk measures, when they lead to fundamentally different decisions.

Unlike risk, reliability theory, engineering, and related methods are not commonly derived from the tuple $(\mathcal{S},L,C)$. However, we argue that they should and that all reliability measures rely on common interpretations of $L$ and $C$, forming a single fundamental perspective. Like risk, reliability interprets $L$ to represent probabilities, such that the same likelihood map $P$ is appropriate in both cases and can be identical for both decision contexts. Unlike risk, reliability depends on a definition of system \textit{quality} that inherently aggregates system states with similar consequences to be treated equally, e.g., as failed or operational. The goal of reliability analysis is to estimate the likelihood of entering undesirable or failed states. This means reliability analysis does not measure consequences via a scalar or numeric value, and instead interprets $C$ to a discrete element:

\begin{itemize}
    \item \textit{Consequence Interpretation as a Category, Group, or Discrete Set-value:} To interpret $C$ as ``meeting'' or ``not meeting'' a specified quality involves filtering its numeric value into a set of discrete categories or groups, e.g., $\mathcal{G} = \{high, medium, low, \dots, g_{k} \}$ via point-wise composition  $\Gamma(\mathcal{S})$: $\mathcal{S} \rightarrow \mathcal{G}(g_{k})$. Here, $\mathcal{G}$ is the resulting set of states organized into $2^{|G|}$ categories, e.g., if states $s_1$ is $high$ and $s_2$ is $medium$, then $\Gamma(\mathcal{S})$ yields $\mathcal{G} =\{\{s_{1}\},\{s_{2}\}\}$. 
\end{itemize}

We define reliability as a perspective on $\mathcal{V}$, specifically $Reliability(\mathcal{V}) = (\mathcal{S},P(\mathcal{S}),\Gamma(\mathcal{S}))$, or simply $Reliability = (P,\Gamma)$, where $\Gamma$ stands for group of discrete elements. This definition is generic to any number of settings, but most reliability analysis are simplified to binary categories via a limit state function \citep{MadsenKrenkLind1986,MelchersBeck2018}. For engineering systems, scenarios that do not exceed a certain level of consequences are often considered \textit{operational}, and those that do are considered \textit{non-operational} or \textit{failed}. Given $C$ returns a non-negative real value, we can define the interpretation of binary states with consequence threshold $\tau$,

\begin{equation}
    \Gamma_{\tau}: \mathcal{S} \rightarrow \mathcal{G}(\tau) = 
   \begin{cases} 
       \text{operational}, & C(s)<\tau,\\ \text{non-operational}, & C(s)\ge\tau. 
   \end{cases}
   \label{eq:interpret-reltau}
\end{equation}

The majority of reliability engineering methods use the associated perspective $Reliability = (P,\Gamma_{\tau})$. With this definition, we can now form common metrics for the reliability of an engineering system, including mean time to failure (MTTF), mean time to repair (MTTR), and related stochastic estimates of first crossing into a non-operational state. Unsurprisingly, reliability measures for availability, unavailability, or otherwise that underlie these metrics ignore individual state consequence value and are only dependent on the likelihood value itself, binned by its associated category. Note that these measures can be derived with the exact same probabilities used in the risk perspective. Again, it is not how we define probability of a given state that matters, it is simply how we use it to interpret the consequences of that state.

\subsection{Examining Differences between Risk and Reliability}

We demonstrate the difference between $Risk(\mathcal{V})$ and $Reliability(\mathcal{V})$ through a simple example. This example shows that common risk and reliability measures will draw attention to different failure scenarios and can make conflicting design recommendations for an engineering system. 


\subsubsection{Example Vulnerability Space and Measures}
Let an engineering system have three known states $\mathcal{S}=\{s_{0}, s_{1}, s_{2}\}$, where $s_{0}$ represents a nominal operational state (i.e., $C(s_{0}) = 0$) and $s_{1}$ and $s_{2}$ are failure states with non-negative consequences $C(s_{1}) = a$, $C(s_{2}) = b$, with $a < b$. We assume that the values of consequences of each state are partially ordered and used directly for analysis, i.e., $G(s) = C(s)$ and the likelihood of each state is

\begin{equation}
    L(s_0)=1-p_{1}-p_{2},
    \qquad
    L(s_1)=p_{1},
    \qquad
    L(s_2)=p_{2},
    \qquad0 < p_{2} < p_{1} < 1,
    \notag
\end{equation}


and our interpretation of each likelihood is a probability with these values $P(\mathcal{S}) = p(s)$. We further assume that $s_{1}$ is a frequent, but less impactful event, and $s_{2}$ is rare, but catastrophic, such that $p_2b>p_1a$.

Given this setup, we define a risk perspective $Risk = (P,G)$, where our interpretation of consequences, $G(\mathcal{S})$, returns their current values $0, a,$ and $b$. We define an associated common risk measure on $Risk(\mathcal{V})$ as the expected loss of transition from the operating state $s_{0}$:

\begin{equation}
V^{\text{risk}} =\sum_{s\in\mathcal{S}}p(s) \cdot C(s).
\label{eq:risk}
\end{equation}

We also define a reliability perspective $Reliability = (P,\Gamma_{\tau})$ based on Eq.~\ref{eq:interpret-reltau} to interpret $C(s)$ into operational and non-operational states. For a performance threshold
$\tau>0$, define the failure set

\begin{equation}
\mathcal{F}(\tau) \subseteq \mathcal{G}(\tau) = \{s \in \mathcal S:C(s) \ge \tau\},
\label{eq:failure-set}
\end{equation}

and associated measure of reliability as the total probability of system failure:

\begin{equation}
V^{\text{unreliable}} = \sum_{s \in \mathcal{F}(\tau)}p(s).
\label{eq:reliability}
\end{equation}


Together, Eq.~\ref{eq:risk} gives initial measure of risk for our generic system as $V^{\text{risk}}_{0} = p_{1}a+p_{2}b$. In contrast, if we assume all failure states are non-operational, such that $\tau < a < b$, then an initial measure of reliability with Eq.~\ref{eq:reliability} is $V^{\text{reliable}}_{0} = p_{1} + p_{2}$. We find this small difference in calculation can have large impacts on decisions to manage engineering system vulnerabilities.

\subsubsection{Example Decision Context}

Now assume we have two engineering designs $d \in D = 1,2$ that we can implement to manage system vulnerability. Design $1$ eliminates the consequences of a frequent moderate failure, whereas design $2$ eliminates a rare catastrophic failure. Table~\ref{tab:generic-3state} presents the resulting consequences across designs, where $C_{0}$ are the consequences given no change to the system, $C_{1}$ given design $1$, and $C_{2}$ for design $2$.

\begin{table}[htbp]
\centering
\caption{
3-State Engineering System Vulnerability and Design Context.
}

\renewcommand{\arraystretch}{1.2}
\begin{tabular}{c|l|c|c|c|c}
\hline
\textbf{Scenario} & \textbf{Description} & \textbf{Probability} & \textbf{$C_{0}(s)$} & \textbf{$C_{1}(s)$} & \textbf{$C_{2}(s)$} \\
\hline
$s_{0}$ & Normal operations & $1-p_{1}-p_{2}$ & $0$ & $0$ & $0$\\
$s_{1}$ & Frequent, moderate disruption & $p_{1}$ & $a$ & $0$ & $a$\\
$s_{2}$ & Rare, severe disruption & $p_{2}$ & $b$ & $b$ & $0$\\

\hline
\end{tabular}
\label{tab:generic-3state}
\end{table}

Each vulnerability measure $m \in M = \{\text{risk}, \text{unreliable}\}$ can be used to define a different objective. Ideally, we want to find a design that minimizes all vulnerability measures simultaneously, such that the union set of designs that minimize each vulnerability measure is not empty:

\begin{equation}
D^{*} = \bigcap_{m\in\mathcal M}
\arg\min_{d\in\mathcal D}
V_d^m,~\text{and}~D^{*} \neq \varnothing.
\label{eq:ideal-vulnerability-design}
\end{equation}

But we find for this simple case, we cannot minimize risk and reliability at the same time as the entire design space is efficient (i.e., Pareto optimal). Minimizing for risk optimally chooses design $2$, as $V^{\text{risk}}_2 = p_{1}a$, which is less than $V^{\text{risk}}_1 = p_{2}b$. In contrast, minimizing for reliability optimally chooses design $1$, as $V^{\text{unreliable}}_{1} = p_{2}$, which is less than $V^{\text{unreliable}}_{2} = p_{1}$. This same reversal in optimal designs holds for any $\tau < a$ and any combination of likelihoods and consequences chosen such that

\begin{equation}
b>\frac{p_1}{p_2}a.
\notag
\end{equation}

This decision reversal is relevant for a realistic engineering decision context, such as power grid vulnerability. Here, consequences are measured in MWh of load shed caused by outage scenarios. Our vulnerability measures represent real design objectives, as $V^{\text{risk}}$ is expected unserved energy (EUE) and $V^{\text{unreliable}}$ is loss of load probability (LOLP). 

Table~\ref{tab:powergrid-3state-designs} presents some representative data for a community power system resulting in vulnerability measures in Table~\ref{tab:powergrid-3state-measures}. For the puposes of this example, we assume a $\tau = 10$ MWh, which we denote LOLP$_{10}$. As demonstrated above, design 1 clearly improves reliability via LOLP$_{10}$ more than EUE, where design 2 does the opposite and minimizes risk. This result holds for $0<\tau<15$ MWh.

\begin{table}[htbp]
\centering
\caption{
Example 3-State Power Grid Vulnerability and Design Context.
}

\renewcommand{\arraystretch}{1.2}
\begin{tabular}{c|l|c|c|c|c}
\hline
\textbf{Scenario} & \textbf{Description} & \textbf{Probability} & \textbf{$C_{0}(s)$} & \textbf{$C_{1}(s)$} & \textbf{$C_{2}(s)$} \\
\hline
$s_{0}$ & Normal operations & $0.89$ & $0$ & $0$ & $0$\\
$s_{1}$ & Frequent, moderate blackout & $0.10$ & $15$ & $0$ & $15$\\
$s_{2}$ & Rare, severe blackout & $0.01$ & $1000$ & $1000$ & $0$\\

\hline
\end{tabular}
\label{tab:powergrid-3state-designs}
\end{table}

\begin{table}[htbp]
\centering
\caption{
Expected Unserved Energy and Loss of Load Probability for a 3-State Power System.
}

\renewcommand{\arraystretch}{1.2}
\begin{tabular}{c|c|c}
\hline
\textbf{Design} & 
\textbf{EUE} [MWh] & \textbf{LOLP$_{10}$} [0,1]
\\
\hline
N/A & $11.5$ & $0.11$ \\
Design 1 & $10.0$ & \textbf{0.01} \\
Design 2 & \textbf{1.5} & $0.10$ \\
\hline
\end{tabular}

\label{tab:powergrid-3state-measures}
\end{table}

\subsection{Implications}
This example highlights one of the key propositions in this work --- different perspectives on the same vulnerability data can fundamentally change engineering design decisions. We argue that our mathematical treatment of data comprising a vulnerability space $\mathcal{V}$ and perspectives comprising interpretations of $L$ and $C$ are necessary first steps before any relevant engineering system vulnerability measure can be constructed and used for decision-making. The act of choosing a vulnerability measure imposes a perspective, even if not recognized by the person producing the analysis. 

The specific reversal in decision-making shown here draws attention to several fundamental distinctions between measures and analysis for risk $(P,G)$ and reliability $(P,\Gamma)$. Mathematically, risk and reliability measures will regularly conflict as they preserve different information about state consequences. For this example, filtering all consequences greater than $\tau = 10$ MWh to be treated the same naturally leads to the decision to prioritize frequent outages at the expense of rare, catastrophic events. Shifting the value of $\tau$ to be greater than $15$ MWh would now make both vulnerability measures optimally choose design $2$. In contrast, risk will deprioritize decisions that manage states with low consequences and likelihoods relative to each other. For example, if the consequence of $s_{2}$ was an order of magnitude smaller (i.e., $C_{0}(s_{2}) = 100$), then both reliability and risk would optimally choose design $1$. 

Practically, this means that risk and reliability measures incur a different analytical burden to make a decision, which prioritizes different activities that do not necessarily inform each other. In a large-scale and complex engineering system, there might not be sufficient time or information to estimate all relevant system states, likelihoods, and consequences for decision-making. In this practically infinite and uncountable vulnerability space, analysts and decision-makers must make expert judgment on which states to evaluate $L$ and $C$ and how many evaluations constitute sufficient information to decide between designs. It turns out, choice of perspective influences what defines sufficient information.

When taking a risk perspective, it makes practical sense to reduce the analytical effort by limiting which likelihoods matter, and spend more analytical effort evaluating limited state consequences. Prior to analysis, it may be difficult to estimate the consequences of events with limited data and that may be the result of complex interactions. Many risk metrics provide a predetermined estimate of the least likely event worth considering (e.g., 1 in 100-year storm, 1 in 10-year blackout, and 1 in 10-year boil water warning) to limit the number of states required to evaluate. This means the analytical effort focuses on consequence measurement of system states that meet this likelihood criteria, with an emphasis on states that are likely and consequential enough under this likelihood filter. As shown in our example, if new information revealed shifts in the relative consequences of $s_{1}$ and $s_{2}$, optimal designs changed without changing reliability decisions.

In contrast, a reliability perspective requires that we determine the performance ``boundary'' separating acceptable and unacceptable system behavior induced by $\tau$ and quantifying the likelihood of crossing that boundary. The associated analytical effort focuses on estimating likelihoods around this boundary, rather than measuring consequences. In most engineering systems, the consequence of a scenario monotonically increases with the intensity of the events that produce it, such that worse events cause equal or worse consequences.  This means that once failure states are found on the boundary, further consequence measurement of worse scenarios is not necessary \citep{barrow2019network}. Similarly, it is unnecessary to evaluate the consequences of states that are known to be operational and far from the boundary produced by $\tau$. 


Finally, this example shows how the choice of limits on likelihood and consequence are key parameters embedded in risk and reliability measures. As discussed, minor changes in $\tau$ can lead to entirely different reliability decisions. This means that the choice of $\tau$ should not be arbitrary, as it directly impacts what designs will be chosen. Likewise, if rare events are not considered in risk analysis, then optimal decisions may differ from those that include them. These practical issues are often ignored in the development and implementation of engineering standards and practices.

\section{Adversaries as a Perspective on Vulnerability}

One argument against defining $Risk = (P,G)$ is that this perspective only comprises probabilistic risk analysis (PRA). In the literature, risk management tends to be defined more broadly to include non-probabilistic estimates of likelihood as well. A key mathematics informing a different perspective on likelihood is possibility theory \citep{zadeh1978fuzzy,dubois2023possibility}, which develops values of state possibilities $\pi(s)$ that are fundamentally different in nature and analysis than probabilities $p(s)$. 

\begin{itemize}
    \item \textit{Likelihood Interpretation as Possibility:} To interpret $L$ as a possibility, the map $\Pi$ is an operator on the entire function space $L$ to ensure it meets the basic definitions of a possibility space, i.e.,  $\Pi:$ $[0,1]^{|\mathcal{S}|} \rightarrow \Pi(\mathcal{S})$ where $[0,1]^{\mathcal{S}}$ is the entire set of likelihoods $\{L \mid L:\mathcal{S}\rightarrow[0,1]\}$ and $\Pi(\mathcal{S})$ ensures a proper possibility distribution 
    $\{\pi:\mathcal{S} \rightarrow [0,1] \mid \sup_{s \in \mathcal{S}}{\pi(s) = 1} \}$. Assuming finite $\mathcal{S}$ and $\max_{t \in \mathcal{S}}{L(s)>0}$, then a standard normalization is $\pi(s) = \frac{L(s)}{\max_{t\in \mathcal{S}}{L(t)}}$.
\end{itemize}

We denote the perspective treating likelihoods as possibilities and measuring consequences as natural numbers as $Adversary(\mathcal{V}) = (S, \Pi(S), G(S))$, or simply $Adversary = (\Pi,G)$, with $\Pi$ meaning possibility.

The key difference between $P$ and $\Pi$ is that $\Pi$ ensures the combination and comparison of likelihoods to be idempotent, i.e., they use maxitive uncertainty measures, rather than additive measures. This is best understood in the normalization of relative possibilities which fundamentally changes the interpretation of system vulnerability. Specifically, the probability of no change from state $s$ (i.e., nothing happens) in $P(\mathcal{S})$ is relative to the probabilities of transition to other states, i.e., $1 - \sum_{s \in \mathcal{S} \backslash s}{p(s')}$. In contrast, by definition, in $\Pi(\mathcal{S})$ at least \textit{one state must be possible} or associated vulnerability measures might not produce an output. Unlike risk or reliability, where no failure might be very unlikely ($0 << p(s) \neq 1$), it is natural from an adversarial perspective to interpret the absence of attack or failure as always possible ($\pi(s)=1$), which ensures coherent measures. 

We find this small difference in normalization has wide implications on the measurement of engineering system vulnerability and the recommendation for design.

\subsection{Examining Differences between Adversary, Risk, and Reliability}

We build on our previous example of an engineering system but add and modify data and interpretation to include additional system states and a possibility distribution.

\subsubsection{Example Vulnerability Space and Measures}

 Assume that the rare, catastrophic event is determined to be much less likely, the moderate disruption is not ideal, but within normal operational plans, and there is now an additional failure state $s_{3}$ that is less frequent than $s_{1}$ but more consequential. We assume our previous natural, scaled consequence interpretation as risk yielding $C(s_{3}) = G(s_{3})=c$ for the new state. And $c$ is chosen with likelihood $L(s_{3}) = p_{3}$, such that



\begin{equation}
0<p_{3}<p_1, \qquad a<p_{3}c<p_{1}a, \qquad p_{2}b < p_{1}a,
\qquad
p_1+p_2+p_{3}<1,
\notag
\end{equation}

given a probability interpretation $P$. The inclusion of $s_{3}$ also renormalizes the likelihood of no failure, $s_0$, to now be

\begin{equation}
L(s_0)=1-p_{1}-p_{2}-p_{3}.
    \notag
\end{equation}

With this new setup, we can still measure risk the same as before using Eq.~\ref{eq:risk}, which is now $V^{\text{risk}}_{0}=p_{1}a+p_{2}b+ p_{3}c$ for no change in design. We assume that our reliability measure is now concerned with large and severe disruptions, not moderate disruptions, such that $a<\tau<c$. This yields a new measure of initial unreliability $V^{\text{unreliable}}_{0}=p_{2} + p_{3}$.    

However, $P(\mathcal{S})$ is an inappropriate interpretation for possibility theory as no state is entirely possible (i.e., there is no state with $p(s)=1$). Hence, we develop a new interpretation $\Pi(\mathcal{S})$ that imposes the following plausibility $\pi(s)$ for each state:

\begin{equation}
\pi(s_0)= \frac{p_0}{p_{0}} = 1,
\qquad
\pi(s_1)=\frac{p_1}{p_{0}},
\qquad
\pi(s_2)=\frac{p_2}{p_{0}},
\qquad
\pi(s_3)=\frac{p_3}{p_{0}}.
\notag
\end{equation}

Note that $\sum_{s\in\mathcal{S}}{\pi(s)}>1$, such that additive aggregation is not meaningful for any measure based on $\pi(s)$. Instead, we define a commonly used adversary vulnerability measure as the most consequential (i.e., worst-case) scenario of our state space

\begin{equation}
V^{\text{adversary}}
= \max_{s\in\mathcal S}
C(s).
\label{eq:adversary}
\end{equation}

Given Eq.~\ref{eq:adversary} and consequence values $b>c>a$, an initial estimate of the vulnerability of our engineering system is $V^{\text{Adversary}}_{0}=b$.

\subsubsection{Example Decision Context}

Like before, we introduce engineering designs that manage each potential failure state. We now have three designs, $d \in D = {1,2,3}$ with corresponding consequences $C_{d}$. Again, design 1 and 2 eliminate the likely, moderate scenario and the rare, catastrophic scenario, respectively. Design 3 eliminates the new scenario, $s_{3}$. Table~\ref{tab:generic-4state} presents the full decision context for this example.

\begin{table}[htbp]
\centering
\caption{
Four State Engineering System Vulnerability and Design Context.
}

\renewcommand{\arraystretch}{1.2}
\begin{tabular}{c|l|c|c|c|c|c|c}
\hline
\textbf{Scenario} & \textbf{Description} & \textbf{Probability} &
\textbf{Possibility}&
\textbf{$C_{0}(s)$} & \textbf{$C_{1}(s)$} & \textbf{$C_{2}(s)$} &
\textbf{$C_{3}(s)$}\\
\hline
$s_{0}$ & Normal operations & $1 - p_{1} - p_{2} - p_{3}$ & 1 & $0$ & $0$ & $0$ & $0$\\
$s_{1}$ & Frequent, moderate disruption & $p_{1}$ & $p_{1}/p_{0}$ & $a$ & $0$ & $a$ & $a$\\
$s_{2}$ & Rare, severe disruption & $p_{2}$ & $p_{2} /p_{0}$ & $b$ & $b$ & $0$ & $b$\\
$s_{3}$ & Somewhat likely, large disruption & $p_{3}$ & $p_{3}/p_{0}$ & $c$ & $c$ & $c$ & 0\\

\hline
\end{tabular}
\label{tab:generic-4state}
\end{table}

Like our previous example, we wish to find some design that can minimize vulnerability across multiple measures, such that $D^{*}$ in Eq.~\ref{eq:ideal-vulnerability-design} is not empty. But each measure points to a separate optimal design. In this case, minimizing $V^{\text{risk}}$ chooses design 1, $V^{\text{unreliable}}$ chooses design 3, and $V^{\text{adversary}}$ chooses design 2. 

This is again easily shown using representative data for a power system. Tables~\ref{tab:powergrid-4state-designs} and~\ref{tab:powergrid-4state-measures} present data for a community-scale power system that corresponds to our generic 4-state engineering example. Here, EUE is calculated the same as before, but we set $\tau = 20$, such that we measure reliability as LOLP$_{20}$. Our vulnerability measure $V^{\text{adversary}}$ corresponds to the real design measure focused on the worst-case unserved energy event in our scenario set, we refer to as UE$^{\max}$. Table~\ref{tab:powergrid-4state-measures} shows that optimal choice of design changes for each measure.

\begin{table}[htbp]
\centering
\caption{
Example 4-State Power Grid Vulnerability and Design Context.
}

\renewcommand{\arraystretch}{1.2}
\begin{tabular}{c|l|c|c|c|c|c|c}
\hline
\textbf{Scenario} & \textbf{Description} & \textbf{Probability} & \textbf{Possibility} & \textbf{$C_{0}(s)$} & \textbf{$C_{1}(s)$} & \textbf{$C_{2}(s)$} & \textbf{$C_{3}(s)$} \\
\hline
$s_{0}$ & Normal operations & $0.849$ & $1$ & $0$ & $0$ & $0$ & $0$\\
$s_{1}$ & Frequent, moderate blackout & $0.10$ & $0.12$ & $15$ & $0$ & $15$ & $15$\\
$s_{2}$ & Rare, severe blackout & $0.001$ & $0.0012$ & $1000$ & $1000$ & $0$ & $1000$\\
$s_{3}$ & Somewhat likely, large disruption & $0.05$ & $0.06$ & $20$ & $20$ & $20$ & $0$\\

\hline
\end{tabular}
\label{tab:powergrid-4state-designs}
\end{table}

\begin{table}[htbp]
\centering
\caption{
Risk, Reliability, and Worst-case Measures for a 4-State Power System
}

\renewcommand{\arraystretch}{1.2}
\begin{tabular}{c|c|c|c}
\hline
\textbf{Design} & 
\textbf{EUE} [MWh] & \textbf{LOLP$_{20}$} [0,1] &
\textbf{UE$^{\max}$} [MWh]
\\
\hline
N/A & $3.5$  & $0.051$ & $1000$ \\
Design 1 & \textbf{2}  & $0.051$ & $1000$ \\
Design 2 & $2.5$  & $0.05$ & \textbf{20} \\
Design 3 & $2.5$  & \textbf{0.001} & $1000$ \\
\hline
\end{tabular}

\label{tab:powergrid-4state-measures}
\end{table}

This example also shows the same divergence in selecting an optimal design when considering additional data necessary for more practical decision settings. As discussed above, in large, complex engineering systems with an uncountably infinite state space, practical application of $V^{\text{risk}}$ requires a filter on the likelihoods of states to analyze, such that only events with sufficient likelihood are included (e.g., 100-year storm). We can mathematically represent this for risk as a {\em probability filter}, $\alpha$, where risk is only measured for states that are sufficiently likely $\mathcal{S}^{\alpha}_{P} = \{s \in \mathcal{S} \mid p(s) > \alpha\}$. An associated risk measure would be 

\begin{equation}
    V^{\text{risk}}(\alpha) =\sum_{s\in\mathcal{S}^{\alpha}_{P}}p(s) \cdot C(s).
    \label{eq:risk-alpha}
\end{equation}

$V^{\text{adversary}}$ suffers from the same practical analysis issue as risk because it uses a natural, scaled consequence interpretation. In an uncountable state space, the focus of adversary analysis is finding which state produces the worst-case $C(s)$ from a subset of plausible states. This is why adversary analysis of an engineering system is usually constrained to a failure or attack budget $N-k$ in practical settings, where the value of $k$ defines events that sufficiently plausible. We can define this {\em plausibility filter} for our possibility distribution by using the same $\alpha$, producing the set of states $\mathcal{S}^{\alpha}_{\Pi} = \{s \in \mathcal{S} \mid \pi(s) > \alpha\}$. An associated adversary measure would be

\begin{equation}
V^{\text{adversary}}(\alpha)
= \max_{s\in\mathcal{S}^{\alpha}_{\Pi}}
C(s).
\label{eq:adversary-alpha}
\end{equation}

Table~\ref{tab:powergrid-4state-measures-alpha} presents the new vulnerability measures using data from Table~\ref{tab:powergrid-4state-designs} and $\alpha = 0.001$, with corresponding power system vulnerability measures EUE$_{\alpha}$ via Eq.~\ref{eq:risk-alpha} and UE$_{\alpha}^{\max}$ via Eq.~\ref{eq:adversary-alpha}. While the differences between probability and possibility values in Table~\ref{tab:powergrid-4state-designs} appear insignificant, they actually allow for choice of $\alpha$ that produces different scenario sets for each analysis, i.e., $\mathcal{S}^{\alpha}_{P} \neq \mathcal{S}^{\alpha}_{\Pi}$. Specifically, any $\alpha$ in the range $0 \leq \alpha < p_{2}/p_{0}$, will remove state $s_{2}$ from the risk measure, but not from the adversary measure. Accordingly, adversary measures do not change even though the system appears to have less risk than before. Design recommendations also do not change as ${s_{2}}$ drives decisions under UE$^{\max}_{\alpha}$ and design 1 produces the least risk. Reliability remains the same as $\alpha$ is not considered it its analysis. Together, each measure still recommends entirely different designs.

\begin{table}[htbp]
\centering
\caption{
Risk, Reliability, and Worst-case Measures for a 4-State Power System with $\alpha = 0.001.$
}

\renewcommand{\arraystretch}{1.2}
\begin{tabular}{c|c|c|c}
\hline
\textbf{Design} & 
\textbf{EUE}$_{\alpha}$ [MWh] & \textbf{LOLP$_{20}$} [0,1] &
\textbf{UE$^{\max}_{\alpha}$} [MWh]
\\
\hline
N/A & $2.5$  & $0.051$ & $1000$ \\
Design 1 & \textbf{1}  & $0.051$ & $1000$ \\
Design 2 & $1.5$  & $0.05$ & \textbf{20} \\
Design 3 & $1.5$  & \textbf{0.001} & $1000$ \\
\hline
\end{tabular}

\label{tab:powergrid-4state-measures-alpha}
\end{table}

\subsection{Implications}

Several authors throughout the literature attempt to align and analyze $(P,G)$ and $(\Pi,G)$ perspectives \citep{aven2011risk,aven2011some,aven2014uncertainty,zio2025advances}. These studies tend to treat possibility distributions as interval functions with bounds set by probabilities and apply $\alpha-$thresholding functions that find median-like measures via a Sugeno integral \citep{dubois2013qualitative}. This approach naturally transforms a pure possibility analysis into one that appears closer to risk and affords better comparison. 

But these works are inconsistent from a pure possibility view by applying probability likelihoods first and then aligning possibility with it, filtering possibilistic analysis through a probability lens. When analyzing a possibility distribution directly, it becomes clearer that these perspectives are fundamentally different as measures in $\Pi(\mathcal{S})$ do not align with the axioms of coherent risk measures. For example, fundamental differences in normalization, which is often ignored in many studies linking probability and possibility, means even in situations where an $\alpha$-thresholding metric is applied, both perspectives produce distinct decisions. 

This is a critical issue as $\alpha$ \textit{is chosen prior to analysis}, such that its influence on final decisions is hidden from the analyst. As the intended the purpose of $\alpha$ in most vulnerability measures is to shift analysis focus on different parts of a likelihood distribution (e.g., as in CVaR), its influence on decision-making is only known if the full state space is already enumerated. However, practical system cannot be fully enumerated, and, instead $\alpha$ acts as a filter on the relevant states for analysis, actually shrinking the search space necessary to make a decision. Hence, arbitrary choice of $\alpha$ and only using a risk or adversary measure may blind the analyst to the diverging decisions they just imposed if measures changed. 

We argue this is partially why there has yet to be a resolution to decades long efforts that distinguish between the decisions made based via engineering interdiction models that identify possible worst-case system failures and PRA-related methods that produce expected failures even if weighted towards the tail of the distribution. Interdiction models, also referred to as attacker-defender models, are designed using optimization and game theory to identify worst-case scenarios for an engineering system of $N$ components given a predefined attack budget $k$, i.e., $N-k$ analysis. Like risk measures which are sensitive to the choice of $\alpha$ threshold on event likelihood, the output of an interdiction model is sensitive to the attack budget $k$. For many systems, slight variations in $k$ can identify completely different sets of critical components \citep{brown_analyzing_2005,alderson_assessing_2014,alderson_operational_2015}. 
Numerous studies show that resilience decisions to manage worst-case failures that result from $N-k$ interdiction models are fundamentally different than those that arise from PRA \citep{willis2007guiding,ezell2010probabilistic,brown2010making,brown2011probabilistic}.




\section{Safety as a Perspective on Vulnerability}

A final key difference between $Risk$ and $\mathcal{V}$ is that we can use $\mathcal{V}$ to define entirely new vulnerability analyses that are inherently different from those already discussed and would, again, produce divergent recommendations from risk if applied to the same engineering system. One common area of research that fits into this category is safety analysis.

Safety has received much less quantitative treatment in the literature than risk and reliability \citep{aven2022risk}. The majority of research developing a quantitative measure of safety do so via generic system frameworks, rather than technical mathematical theory. Some well-established examples originate from the human factors and resilience engineering research communities, including Safety-I and Safety-II \citep{hollnagel2015safety,hollnagel2017safety,hollnagel2018safety}, event analysis for systemic teamwork (EAST) and related frameworks \citep{GibsonEtAl2005EAST,WalkerEtAl2006EAST}, and system-theoretic accident model and prevention \citep{leveson2016engineering,patriarca2022past}, Accident Map (AcciMap) \citep{Rasmussen1997,SvedungRasmussen2002}, and functional analysis resonance method (FRAM) \citep{hollnagel2004functional,patriarca2017defining,patriarca2020framing}. This approach has led to the application of risk, reliability, and adversary mathematical techniques in safety analysis, which means it will not produce novel recommendations distinct from the measures already presented. This often leads to conflating concepts like reliability and safety, despite many examples of reliable systems being unsafe and safe systems being unreliable \citep{leveson2016engineering}. Moreover, they tend to embed significant qualitative information and normalization that prevents broader mathematical and theoretical development outside of human factors and specific engineering contexts.

Using our theory of vulnerability, we now have a simple, straightforward way to define quantitative safety analysis as a perspective on $\mathcal{V}$. Conceptually, safety aligns with a reliability interpretation of consequences, as it tends to focus attention on preventing or avoiding undue harm by ensuring engineering systems they do not enter unsafe operating states. This definition inherently imposes a set-valued perspective on $C$ like $\Gamma$ where states can be categorized as safe and unsafe. Safety also aligns with adveresary interpretation of likelihoods, as it tends to focus design on reducing the possibility of transition to unsafe states, rather than the probability. Accordingly, the most generic safety perspective is $(\Pi,\Gamma)$. Assuming similar treatment of consequences via a limit state function on $\tau$, we define $Safety = (\Pi,\Gamma_{\tau})$.

With this definition, several well-established technical methods for engineering vulnerability analysis naturally align with safety more than other perspectives. These include network survivability analysis, that attempts to measure the maximum number for $k$ that a system can experience before entering a non-functioning state. Historically, survivability has been considered a risk or reliability measure and has been applied in telecommunications systems. In contrast, we argue this and related reliability measures common in utility industries, such as $N-1$ reliability for power systems, should actually be considered the same and a safety measure instead. Similarly, reachability analysis for engineering systems uses a similar framing to estimate the likelihood a system may or may not cross a predefined operational threshold. We argue that any reachability analysis that uses discrete or possibilistic factors for scenario likelihood should be considered safety, and not reliability.

\subsection{Examining Differences between All Four Perspectives}

\subsection{Safety of the 4-State System}
Given a safety measure integrates possibility and set-valued perspectives on $L$ and $C$, we can define a safety measure based on previously defined mathematics. For example, a standard measure of safety relevant for our previous examples would be similar to reliability, but informed by our possibility distribution. Namely, the maximum possibility of crossing the performance threshold $\tau$

\begin{equation}
V^{\text{unsafe}} = \max_{s \in \mathcal{F}(\tau)}\{\pi(s)\},
\label{eq:safety}
\end{equation}

where $\mathcal{F}(\tau)$ is the failure set from Eq.~\ref{eq:failure-set}. 

Using $V^{\text{unsafe}}$ on our previous 4-state example, we find that safety and reliability would both recommend the same optimal decision:  choose design 3. However, the difference in normalization of $P$ and $\Pi$ induce different vulnerability values, such that $V^{\text{unreliable}} = 0.05$ and $V^{\text{unsafe}} = 0.06$ given no design choice, and $V^{\text{unreliable}} = 0.001$ and $V^{\text{unsafe}} = 0.0012$ under design 3. This means that the system appears less safe than reliable, all else equal, and both measures will always recommend the same design given both measures use the same $\tau$ and $\pi(s)$ is derived from $p(s)$. 

In practical decision settings, these assumptions do not normally hold. In general, designing a system to be reliable tends to focus on small values of $\tau$ that correspond with small consequences and indicate continued system operations. In contrast, safety is generally less concerned with maintaining normal operations and is very concerned with safety-critical situations that incur large harms on the system, people, or society. Increasing $\tau$ to reflect this difference in our 4-state system would shift safety from aligning with reliability and recommending design 3 to aligning with adversary and recommending design 2.

Further, the possibility of a state and the plausibility of scenarios often do not conform to the same distribution as their probabilities. For example, the probability of two components failing simultaneously might be vanishingly small in an engineering system, yet such a scenario would be entirely possible in a simple $N-2$ adversary analysis. This implies that $\Pi$ might still conform to similar rules to $P$, such as events with greater impact should associate with monotonically increasing consequences, but need not use the same distribution of values 

\subsection{5-State Power Grid Vulnerability Space and Decision Context}

Now assume we now have a five state vulnerability space for a community power system with four designs that can manage its vulnerability. In this case, the probabilities and possibilities of each state are determined via independent analysis. Table~\ref{tab:powergrid-5state-designs} presents the resulting probability, possibility, and consequences associated with each state and design.

\begin{table}[htbp]
\centering
\caption{
Example 4-State Power Grid Vulnerability and Design Context.
}

\renewcommand{\arraystretch}{1.2}
\begin{tabular}{c|c|c|c|c|c|c|c}
\hline
\textbf{Scenario} & \textbf{Probability} & \textbf{Possibility} & \textbf{$C_{0}(s)$} & \textbf{$C_{1}(s)$} & \textbf{$C_{2}(s)$} & \textbf{$C_{3}(s)$} &
\textbf{$C_{4}(s)$}
\\
\hline
$s_{0}$ & $0.839$ & $1$ & $0$ & $0$ & $0$ & $0$ & $0$\\
$s_{1}$ & $0.07$ & $0.75$ & $15$ & $0$ & $15$ & $15$ & $15$\\
$s_{2}$ & $0.001$ & $0.01$ & $1000$ & $1000$ & $0$ & $1000$ & $1000$\\
$s_{3}$ & $0.05$ & $0.25$ & $20$ & $20$ & $20$ & $0$ & $20$\\
$s_{4}$ & $0.04$ & $0.5$ & $25$ & $25$ & $25$ & $25$ & $0$\\

\hline
\end{tabular}
\label{tab:powergrid-5state-designs}
\end{table}

Note the values in each distribution align in some ways, but are different, which changes the order of which states are more or less likely. Specifically, $s_{0}$ is the most likely state under each interpretation and state $s_{2}$ is the least. However, states $s_{1}$, $s_{3}$ , $s_{4}$ are determined to have similar, but different probabilities, and each state has a distinct possibility. This could correspond to minor differences in the detailed calculation of the probability of the event dictating one failure state is more or less likely (e.g., due to age, location, etc.), where possibilities might align with expert knowledge on what events would be more or less plausible (e.g., $s_{1}$ appears more possible than $s_{3}$ or $s_{4}$).

Table~\ref{tab:powergrid-5state-measures-alpha} presents the result of each design given our vulnerability metrics EUE, UE$^{\max}$, and LOLP$_{20}$ for risk, adversary, and reliability. We also define the safety of the system as the loss of load possibility, LOL$\Pi_{20}$ via Eq.~\ref{eq:safety}, which can be interpreted as the reachability of crossing our performance threshold $\tau$. 

\begin{table}[htbp]
\centering
\caption{
Vulnerability Measures for a 5-State Power System}

\renewcommand{\arraystretch}{1.2}
\begin{tabular}{c|c|c|c|c}
\hline
\textbf{Design} & 
\textbf{EUE} [MWh] & \textbf{LOLP$_{20}$} [0,1] &
\textbf{UE$^{\max}$} [MWh] &
\textbf{LOL$\Pi_{20}$} [0,1] 
\\
\hline
N/A & $4.1$  & $0.091$ & $1000$ & 0.5 \\
Design 1 & \textbf{3}  & $0.091$ & $1000$ & 0.5 \\
Design 2 & $3.1$  & $0.09$ & \textbf{25} & 0.5 \\
Design 3 & $3.1$  & \textbf{0.041} & $1000$ & 0.5 \\
Design 4 & $3.1$  & $0.051$ & $1000$ & \textbf{0.25} \\
\hline
\end{tabular}
\caption*{\textit{\textbf{Note:}} The same decisions hold for EUE$_{\alpha}$ and UE$^{\max}_{\alpha}$ with  $0 \leq \alpha <0.01$.}

\label{tab:powergrid-5state-measures-alpha}
\end{table}

With our final example, we show that realistic choices for data values that comprise vulnerability measures optimally choose different designs. Here, risk chooses design 1, adversary design 2, reliability design 3, and safety design 4. However, we do find some potential efficiencies among design choices. For example, design 4 is not the optimal choice to minimize $V^{\text{unreliable}}$, but is preferred over designs 1 and 2. Design 1 is equally bad from a reliability perspective as doing nothing. In contrast, risk is reduced given any design, but only design 1 achieves minimum risk. From a possibility perspective, only design 2 benefits adversary measures and only design 4 benefits safety. This means there are significantly more equivalent solutions that are equally bad and difficult to distinguish using these measures.




\section{Towards a Theoretical Framework for Quantitative Vulnerability Analysis}

Figure~\ref{fig:vulnerability-decision-spaces} presents the unified framework showing how perspectives on $\mathcal{V}$ produces risk, reliability, adversary, and safety analyses. We further provide examples of several measures commonly used for vulnerability analysis of engineering systems and organize them in the perspectives they align. The measures defined in this work used to demonstrate differences in computation, results, and decisions only constitute a small subset of the measures used in practice. We discuss the implications of this new theoretical framework with respect to how it supports new theory for system resilience, interpretation of more complex, hybrid vulnerability measures, and multiobjective optimization for integration of perspectives.

\subsection{Formal Conceptual Reframing}

A fundamental result of our framework is that there are at least four vulnerability analysis perspectives that should be maintained as distinct and should not be parent to, or conflated with, each other. Table~\ref{tab:risk-reliability-conceptual-errors} summarizes some of the conceptual errors our framework overcomes for risk and reliability as an example of conceptual errors across all four forms of analysis. Our framework overcomes these errors by linking the mathematical processes of normalization via a measure and the preference ordering via a quasimetric to their practical use in system analysis and design decisions. This result means that system data cannot be used directly in decisions (i.e., $\text{Data} \mathrel{\nRightarrow} \text{Decisions}$), and must be interpreted prior to forming a decision model (i.e, $\text{Data} \mathrel{\Rightarrow} \text{Perspective} \mathrel{\Rightarrow} \text{Decisions}$). This further suggests that a given perspective like $(P,G)$ (risk) and associated decisions using this perspective should be considered entirely separate from counterpart analysis of $(P,\Gamma)$ (reliability), $(\Pi,G)$ (adversary), and $(\Pi,\Gamma)$ (safety). This is in contrast to research that re-arrange this framing, or consider one concept or analysis interchangeable for another in terms of decision utility and treatment of data.

\begin{table}[htbp]
\caption{Conceptual Errors in Relating Risk and Reliability.}
\label{tab:risk-reliability-conceptual-errors}
\begin{tabular}{p{0.20\textwidth} p{0.35\textwidth} p{0.37\textwidth}}
\toprule
\textbf{Conceptual Error}
&
\textbf{Common Manifestation}
&
\textbf{Our Framework}
\\
\midrule
\textbf{Conflation}
&
Risk and reliability are ``the same thing.''
&
We show they apply different state-ordering and aggregation rules.
\\
\addlinespace
\textbf{Subsumption}
&
Reliability is a special case of risk, or risk is a special case of
reliability.
&
Neither perspective generally derives from the other nor does it produce the same decision preferences as the other.
\\
\addlinespace
\textbf{Substitution}
&
A reliability measure is treated as a risk measure, or vice versa.
&
Related numerical scores or ordering under one perspective need not preserve rankings under the other.
\\
\addlinespace
\textbf{Reification}
&
Risk and reliability are treated as intrinsic properties of the system
alone.
&
They are perspective-dependent evaluations of system data on state likelihood and consequences. They are not inherent system properties independent of an evaluative perspective.
\\
\bottomrule
\end{tabular}
\caption*{\textit{Note:} We use risk and reliability for example, but adversary and safety perspectives can be interchanged with both.}
\end{table}

Another result of this work is the implication that risk and risk analysis should be reframed as a form of vulnerability analysis, like reliability, rather than as the parent class of concepts that subsumes the others. Some authors will object to the reframing presented here, suggesting that our vulnerability space space should be termed ``risk'' or a ``risk space'' with four sub-types: probabilistic risk $(P.G)$, adversary risk $(\Pi.G)$, reliability risk $(P.\Gamma)$, and safety risk $(\Pi,\Gamma)$. Doing so would align with much of the conceptual literature subsequent to \cite{kaplan1981quantitative}, such as recent development by \citet{Aven2026}, which uses a superficially similar consequence-uncertainty framing, but concludes methods like reliability and safety as forms of risk.



\begin{landscape}
\begin{figure}[htpb]
\centering

\begin{tikzpicture}[
    x=1cm,
    y=1cm,
    axis/.style={line width=1pt, black},
    axislabel/.style={font=\bfseries\normalsize, align=center},
    axissubtitle/.style={font=\itshape\small, align=center},
    quadrant/.style={font=\bfseries\large, align=center},
    formula/.style={font=\normalsize, align=center},
    measures/.style={
        font=\small,
        align=left,
        text width=7.35cm,
        inner sep=0pt,
        outer sep=0pt
    }
]

\def\xmin{0}
\def\xmid{8.35}
\def\xmax{16.70}
\def\ymin{2}
\def\ymid{7.40}
\def\ymax{12.40}

\node[
    font=\bfseries\normalsize,
    align=center,
    rotate=0,
    anchor=south
] at (\xmid,13.00)
{Consequence \\Interpretation};

\node[axislabel] at (4.175,\ymax+0.38)
{Natural or Scaled-value};

\node[axissubtitle] at (4.175,\ymax-0.15)
{\(\displaystyle G(\mathcal{S)}:C(s)\to\mathbb{R}\)};

\node[axislabel] at (12.525,\ymax+0.38)
{Set-value or Categorical};

\node[axissubtitle] at (12.525,\ymax-0.15)
{\(\displaystyle \Gamma_{\tau}(\mathcal{S}) :C(s) \rightarrow \mathcal{G}(\tau)\)};

\node[
    font=\bfseries\normalsize,
    align=center,
    rotate=0,
    anchor=south
] at (-1.5,\ymid-.5)
{Likelihood \\ Interpretation};

\node[axislabel, anchor=east] at (-0.18,10.55)
{Probability};

\node[axissubtitle, anchor=east] at (-0.18,10)
{\(\displaystyle P(\mathcal{S}):L(s) \rightarrow p(s)\)};

\node[axislabel, anchor=east] at (-0.18,5.05)
{Possibility};

\node[axissubtitle, anchor=east] at (-0.18,4.50)
{\(\displaystyle \Pi(\mathcal{S}): L(s) \rightarrow \pi(s)\)};

\draw[axis] (\xmin,\ymid) -- (\xmax,\ymid);
\draw[axis] (\xmid,\ymin) -- (\xmid,\ymax);

\node[quadrant] at (4.175,11.25) {Risk};

\node[formula] at (4.175,10.58)
{\(\displaystyle
\text{Risk}=\bigl(P,G\bigr)
\)};

\node[measures, anchor=north west] at (0.48,10.05) {
\textbf{Representative measures:}\\[2pt]
\(\bullet\) Expected loss or expected consequence\\
\(\bullet\) Value-at-Risk (VaR)\\
\(\bullet\) Conditional Value-at-Risk (CVaR) / expected shortfall\\
\(\bullet\) Expected unserved energy or expected outage cost
};

\node[quadrant] at (12.525,11.25) {Reliability};

\node[formula] at (12.525,10.58)
{\(\displaystyle
\text{Reliability}
=
\bigl(P,\,\Gamma\bigr)
\)};

\node[measures, anchor=north west] at (8.83,10.05) {
\textbf{Representative measures:}\\[2pt]
\(\bullet\) Mean time to failure (MTTF)\\
\(\bullet\) Availability or probability of service\\
\(\bullet\) Loss-of-load probability (LOLP)\\
\(\bullet\) SAIDI and related interruption indices
};

\node[quadrant] at (4.175,6.25) {Adversary};

\node[formula] at (4.175,5.58)
{\(\displaystyle
\text{Adversary}
\bigl(\Pi,\,G\bigr)
\)};

\node[measures, anchor=north west] at (0.48,5.05) {
\textbf{Representative measures:}\\[2pt]
\(\bullet\) Network interdiction\\
\(\bullet\) Maximum load shed under failure budget\\
\(\bullet\) Worst-case \(N-k\) contingency performance
};

\node[quadrant] at (12.525,6.25) {Safety};

\node[formula] at (12.525,5.58)
{\(\displaystyle
\operatorname{Safety}
=
\bigl(\Pi,\Gamma \bigr)
\)};

\node[measures, anchor=north west] at (8.83,5.05) {
\textbf{Representative measures:}\\[2pt]
\(\bullet\) $N-1$ security criterion\\
\(\bullet\) Reserve margin\\
\(\bullet\) Reachability analysis\\
\(\bullet\) Minimum contingency budget to violation
};

\end{tikzpicture}

\caption{
Unified engineering system vulnerability analysis framework and representative quantitative measures.}
\label{fig:vulnerability-decision-spaces}

\end{figure}
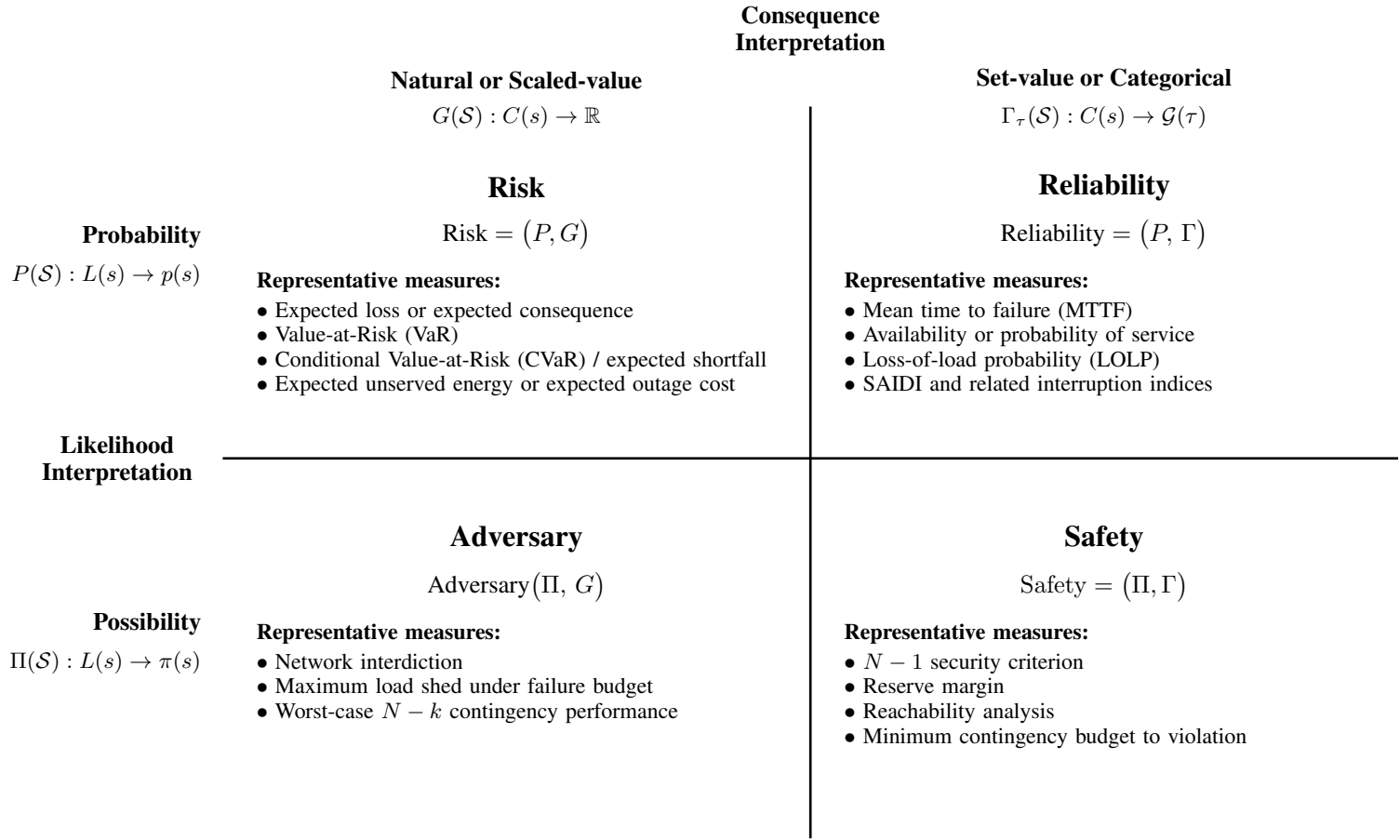
\end{landscape}

However, such a framing would also confuse and conflate the other perspectives with essentially all quantitative methods proposed for risk analysis. Most authors in the literature that made contributions to the quantitative measurement of risk for decision-making do not do so via a vulnerability space of data, but exclusively through the risk perspective proposed here. Important examples include  \cite{kaplan1981quantitative}, as the authors implicitly jump to the $(P,G)$ perspective for risk analysis in the immediate next paragraph after introducing the broader risk triplet.  \citet{aven2011risk} defines the risk triplet as containing probabilities upfront when attempting to compare and integrate risk concepts with possibility theory. The resulting proposed methods implicitly filter possibility distributions through a probabilistic lens, ignoring the flexibility and relationships possibilities also have for adversary and safety analysis presented here. Furthermore, conceptual developments by \citet{zio2025advances} that compare reliability and risk emphasize the definition the $(P,G)$ definition of risk to compare the concepts and their outputs. Hence, our advocacy of treating system vulnerability as the core concept, rather than risk, is deliberate with the intent to ensure risk analysis as a method is appropriately defined alongside its co-constituted and distinct methods of reliability, adversary, and safety analysis.


\subsection{Relationships to Engineering System Resilience}

Redefining risk as a perspective on vulnerability enables us to now provide technical, quantitative rigor to what constitutes each form of analysis. This substantially helps to distinguish between different decisions regarding the design and resilience of the engineering system. Assuming engineering decisions intended to improve resilience should be informed by a vulnerability analysis, our methods clarify how the choice of the same resilience action may or may not be recommended with changes in perspective. Moreover, fundamental differences in the focus, analysis, and results generated by interpretation of $L$ and $C$ suggest that choice of measure within a given perspective may be arbitrary --- making resilience decisions using measures given the same risk perspective, even if recommendations are different, may not necessarily be making different decisions. 


A second, subtle implication of this framework comes from what we did not define nor address at any point in its exposition---resilience. In our view, system vulnerability and resilience are distinct, interacting concepts that relate and interact with each other, but one does not subsume the other, or vice versa (c.f., \citealt{cutter2008place,HollnagelEtAl2019EngineeringResilience}). Philosophically, we treat vulnerability and resilience as mutually constitutive opposites from Hagel's dialectics --- two opposing ideas that interact and evolve through their conflict, leading to a higher level of understanding \citep{Hegel2010EncyclopaediaLogic,Hegel2010ScienceOfLogic}. Practically, this means, vulnerability analysis is a distinct activity that can be performed without yet defining what techniques, actions, capabilities, or otherwise one seeks to improve system resilience. Yet, vulnerability measures on their own are not sufficient, designs and actions that constitute resilience are needed to make an engineering decision. It is the very nature of the vulnerability of engineering systems that allows resilience to exist, and vice versa.

This perspective contrasts with significant research research that compare resilience to risk \citep{park2013integrating,eisenberg2018think,aven2019call,linkov2019science}. In our view, risk is a sub-concept to vulnerability, an operationalization of the concept. Comparing risk to resilience is a categorical mistake as these concepts reside on different conceptual levels. We argue this conceptual mistake that pervades the literature is perhaps one reason that many engineering systems remain vulnerable depicted decades of research \citep{eisenberg2025rebound,alderson2019overcoming,EisenbergAlderson2024OperationalEnergyVR}.

\subsection{Classifying Hybrid Vulnerability Measures and Models}

A key challenge for this framework is the classification of hybrid measures that appear to blend perspectives. We believe classifying measures and decisions represents an important area of research, both to determine what decisions more complex measures are actually supporting and to potentially disprove this theory. However, upon initial inspection many common measures and models that appear to blend decisions might actually embed perspectives in such a way to clearly fit in one over another.  

Some representative examples of hybrid measures that do this include value-at-risk ($VaR$) and derivative measures (e.g., cumulative value at risk $CVaR$, entropic value at risk $EVaR$, buffered probability of exceedance $PBOE$, etc. \citealt{norton2021calculating,chaudhuri2022certifiable}) which apply an $\alpha$-thresholding statistic on the loss distribution of a system. Hence, these measures assume full enumeration of state probabilities and consequences prior to thresholding results, which we established is practically infeasible in complex engineering systems. In systems where such data exists, such measures may appear somewhat like reliability by embedding a threshold. However, this threshold does not bin states into specific categories, it simply shifts the aggregation of likelihoods and consequences from minimizing expected consequences to consider tail-risk that might be less likely, but more consequential. 

$VaR$-like measures might also appear to support both risk and adversary decisions, as $\alpha$ acts as a tuning parameter --- its bounds recover the mean of the loss distribution and the worst-case consequences. However, the resulting worst-case state is based on a probability interpretation of likelihood, which we show does not necessarily align to possibility given differences in normalization. Moreover, a possibility distribution affords assigning $\pi(s)$ to events with no data or prior knowledge, which would produce spurious probabilities can produce meaningless results for a risk measure.

These examples are only a few of those found in the literature, and our analysis here is neither formal nor complete. Future work should focus not only on the development of these hybrid measures, but also on their formal classification as to what decisions they inform. A key question is whether they fundamentally change a design decision or not. And whether interpretation of $\mathcal{S}$ is an important facet to the framework that is currently ignored in this presentation.


\subsection{Multiobjective Decision Analysis}

Given our framework, vulnerability analysis of an engineering system now appears to be a multiobjective decision problem, rather than one that can rely on a single metric. Indeed, developing analysis that can effectively integrate risk, reliability, adversary, and safety together may provide win-win designs that do not suffer from the decision reversals presented in this work.

But an important implication of this work is that techniques to try and integrate decisions across perspectives, may not necessarily work to alleviate the narrowing effect of taking a perspective. Our simple example demonstrates that with relevant, realistic data, it is possible to generate a decision context where the entire feasible space of designs is efficient (i.e., Pareto Optimal). This means every design is non-dominated one of our fundamental metrics, and decision-making comes down to tradeoffs incurred by choosing any design at all. To make more comprehensive analysis, you should at least consider measures from distinct perspectives. But doing so does not make decision-making easier, nor likely to find win-win solution that address vulnerabilities from multiple perspectives. Finding the provable limits of each perspective and their integration is a worthy area of research.




\section{Conclusion}

Engineering systems management requires the use of quantitative measures to identify efficient and optimal designs. Although risk analysis remains a standard approach to prioritize one design over another, its broad definition in the literature does not match its mathematical treatment for decision-making. We believe this lack of rigor has led many researchers and practitioners to use terms like risk, reliability, adversary, safety and their derivatives almost interchangeably. This is further exacerbated by conflated but distinct concepts like resilience and robustness \citep{park2013integrating,aven2019call,logan2022risk}. Without a technical basis for their distinction, we are limited to rely on linguistic and conceptual differences demarcating one over another which engenders a seemingly Sisyphusian effort to integrate and better define terms \citep{mentges2023resilience}. Hence, it may be most pragmatic for an analyst or practitioner to simply choose the measure they deem most appropriate for the engineering system, call it one of these terms, and proceed with decision-making.

Yet, such loose treatment of such important concepts does not come without consequence. The decisions risk, reliability, adversary, and safety analysis make often mean life and death. In the best case, when one says they are making risk-based or risk-informed decisions when they are not, the misrepresentation of their analysis does not change the final design nor harm anyone. In the worst case, they may believe they are reducing risk, when they have used measures and made decisions that have no impact on the risk of the system. This is especially problematic in engineering organizations and sectors that have oriented their entire vulnerability management programs around a single perspective, either implicitly or explicitly, suggesting it is {\it the} way of doing analysis.  This work argues that \textit{there is no single correct perspective,} and organizations that rely on stale or limited methods should reconsider their efforts to include multiple quantitative perspectives.


Our unified quantitative framework is an important start for disaggregating vulnerability analysis methods, but it is likely not the true nor final approach. One reason this unification is important is not necessarily because it is right, but because it can stimulate new ideas. Setting hard and fast rules for what terms and analysis means enables experts to better explain what is important and novel about future measures, models, and decision contexts. This creates constraints that can de-constrain \citep{alderson2010contrasting,doyle2011architecture} the risk analysis community and hopefully enables advances in risk science that may have previously been hampered.

\bibliographystyle{abbrvnat}
\bibliography{refs}  

\end{document}